\documentclass[12pt]{article}

\usepackage{natbib}
\usepackage{apalike}
\usepackage{geometry}
\usepackage[skip=0.5\baselineskip, indent=1.5em]{parskip}

\usepackage{placeins}
\usepackage[percent]{overpic}

\usepackage[utf8]{inputenc} % allow utf-8 input
\usepackage[T1]{fontenc}    % use 8-bit T1 fonts
\usepackage{hyperref}       % hyperlinks
\usepackage{url}            % simple URL typesetting
\usepackage{booktabs}       % professional-quality tables
\usepackage{amsfonts}       % blackboard math symbols
\usepackage{nicefrac}       % compact symbols for 1/2, etc.
\usepackage{xcolor}         % colors
\usepackage{subcaption}

\usepackage{amsmath}
\usepackage{amsthm}
\usepackage{amssymb}
\usepackage{accents}
\usepackage{algorithm,algorithmic}
\usepackage{graphicx}

\usepackage{authblk}

\providecommand{\keywords}[1]
{
  \small	
  \textbf{\textit{Keywords---}} #1
}

\newcommand{\R}{\mathbb{R}}
\let\P\relax 
\DeclareMathOperator{\P}{pr}

\newcommand{\N}{\mathbb{N}}

\newcommand{\Ipq}{I_{p,q}}

\renewcommand{\O}{\mathbb{O}}
\newcommand{\+}[1]{#1}

\renewcommand{\*}[1]{\mathcal{#1}}

\DeclareMathOperator{\diag}{diag}
\DeclareMathOperator{\rank}{rank}

\newcommand{\leqc}{\lesssim}
\newcommand{\leqcP}{\leqc_{\P}}
\newcommand{\geqc}{\gtrsim}

\newcommand{\eqc}{\asymp}

\newcommand{\norm}[1]{\left\| #1 \right\|}
\newcommand{\smallnorm}[1]{\| #1 \|}

\newcommand{\inner}[1]{\left\langle #1 \right\rangle}

\newcommand{\brackets}[1]{\left( #1 \right)}

\newcommand{\curlybrackets}[1]{\left\{ #1 \right\}}
\newcommand{\smallcurlybrackets}[1]{\{ #1 \}}

\newtheorem{theorem}{Theorem}
\newtheorem{proposition}[theorem]{Proposition}

\newtheorem{lemma}[theorem]{Lemma}

\theoremstyle{definition}
\newtheorem{remark}{Remark}
\newtheorem{definition}[theorem]{Definition}

\title{{\bf Spectral embedding and the latent geometry of multipartite networks}}

\author[1]{Alexander Modell\footnote{Corresponding author, \texttt{a.modell@imperial.ac.uk}.}}
\author[2]{Ian Gallagher}
\author[3]{Joshua Cape}
\author[4]{Patrick Rubin-Delanchy}
\affil[1]{Imperial College London, U.K.}
\affil[2]{University of Melbourne, Australia}
\affil[3]{University of Wisconsin–Madison, U.S.A.}
\affil[4]{University of Edinburgh, U.K.}
\date{}

\begin{document}

\maketitle

\begin{abstract}
  Spectral embedding finds vector representations of the nodes of a network, based on the eigenvectors of a properly constructed matrix, and has found applications throughout science and technology. Many networks are multipartite, meaning that they contain nodes of fundamentally different types, e.g. drugs, diseases and proteins, and edges are only observed between nodes of different types. When the network is multipartite, this paper demonstrates that the node representations obtained via spectral embedding lie near type-specific low-dimensional subspaces of a higher-dimensional ambient space. For this reason we propose a follow-on step after spectral embedding, to recover node representations in their intrinsic rather than ambient dimension, proving uniform consistency under a low-rank, inhomogeneous random graph model. We demonstrate the performance of our procedure on a large 6-partite biomedical network relevant for drug discovery.
\end{abstract}


\section{Introduction}
Graph embedding describes a family of tools for representing the nodes of a graph (or network) as points in space. Applications include exploratory analyses such as clustering \citep{ng2002spectral,von2007tutorial} or visualisation \citep{jacomy2014forceatlas2}, predictive tasks such as classification \citep{tang2013universally}  or link prediction \citep{kazemi2018simple}. The purpose of this article is to develop statistical methodology for the case when the graph is multipartite, meaning that the nodes are of known types, and edges only occur between pairs of nodes of different types.

A rigorous statistical treatment of this case would seem important given the wide range of applications in which such graphs are encountered. These include security applications, for example, data connecting users, computers and processes in cyber-security \citep{akent-2015-enterprise-data}; or images, phone-numbers, locations and names in human-trafficking prevention \citep{szekely2015building} and biomedical applications, e.g. data connecting drugs, diseases, targets, pathways, variant locations and haplotypes \citep{zong2021drug}, a 6-partite example analysed in this paper. More broadly, multipartite graphs provide a useful (partial) representation of relational databases. A common construction is to infer node types from the column names, assign a node to each unique entry within a column, and place an edge between two nodes if the corresponding entries occur together in a row. Data scientists employ such constructions to explore large databases using graph analytics and visualisation tools.

% it is very common in data science to construct a multipartite graph from a relational database, to allow interactive data exploration using graph-based tools. A common construction is to infer node types from the column names, assign a node to each unique entry within a column, and place an edge between two nodes if the corresponding entries occur together in a row.

One might have expected that the now substantial body of work on graph modelling by the Statistics community would easily transfer to multipartite graphs, but that is not the case. First, assumptions of transitivity, homophily, or assortativity still seem to dominate the field \citep{grover2016node2vec,athreya2017statistical,todeschini2020exchangeable,crane2021root}. The details differ, but at a high level the assumption is that ``similar entities should connect''. This is a nonsensical starting point for multipartite graphs: it would often be entirely meaningless to say that two entities of different types (e.g. a user and a computer) were ``similar'', and yet that is where the connections occur. Second, knowledge of the node types, which we almost always have in statistical applications, gives us very specific structural information about the graph. In fact, if we did not know the node types, it would be an NP-complete problem just to verify that the graph was multipartite \citep{garey1990computers}. So, how can we exploit this information?

% , which must be exploited. If we did not know the node types, it would be an NP-complete problem just to verify that the graph was multipartite. It will turn out that we can use type information in a much more targeted way that we could, say, generic node labels.

Spectral embedding provides an answer. The central observation of this paper is that the spectral embedding of a multipartite graph has special geometric structure, in which the node representations of each type lie in the vicinity of a type-specific, lower-dimensional subspace (see Figure~\ref{fig:chung_lu}). We can exploit this geometry via a secondary dimension reduction step, which takes the node representations from a global, ambient dimension into a type-specific, intrinsic dimension. The embedding procedure is highly scalable: the first, more compute-critical spectral decomposition, which needs to be applied to the entire graph, is one where we can exploit sparsity \citep{baglama2005augmented}, and the subsequent dimension reduction step can be computed quickly using modern numerical linear algebra procedures \citep{halko2011finding}. Inserting this secondary step into spectral clustering yields a new variant of the algorithm, and the estimated dimension of each subspace indicates how many clusters to select for that node type. 

We establish a strong theoretical guarantee on the resulting embedding --- uniform consistency --- which is justified by the practical context. Embeddings support a vast number of applications, many unforeseen, and so a weaker guarantee could have unforeseen consequences. It is a simple implication of our theory that our proposed variant of spectral clustering produces an asymptotically exact partition of the nodes under a multipartite  stochastic block model.

\subsection{Notation}
 We say a sequence of events $\smallcurlybrackets{E_n}$ occurs \emph{with overwhelming probability} if for any constant $c>0$ there exists $n_0 \in \N$ and $C>0$ (which may depend on $c$, but not $n_0$) such that for all $n \geq n_0$, $\P(E_n) \geq 1 - C n^{-c}$.
For two scalars $a,b$, we write $a \leqc b$ to mean $a \leq C b$ where $C$ is a universal constant which, when qualified with the prior probabilistic statement, may depend on the constant $c$. We write $a \eqc b$ if $a \leqc b$ and $a \geqc b$. For any positive integer $n$, we use the shorthand $[n] := \{1,\ldots,n\}$ and for a matrix $M$, we write $\sigma_i(M)$ to denote its $i$th largest singular value. For any number $s$, $\pm s$ denotes the set $\{-s, s\}$.

\section{Preliminaries}
We first give some background on adjacency spectral embedding for generic, undirected graphs; biadjacency spectral embedding for bipartite graphs; and on the generalised random dot product graph model, which will form the foundation of our geometric interpretation of spectral embedding of multipartite graphs.
% For simplicity, we focus on adjacency spectral embedding however the ideas and theory presented extend to Laplacian spectral embedding and its variants with appropriate modifications \citep{tang2018limit}.

\subsection{Spectral embedding for generic, undirected graphs}

In this paper, a simple, undirected graph with vertex set $\*V := [n]$ is represented by its symmetric adjacency matrix $A \in \{0,1\}^{n\times n}$ with entries $a_{ij} = 1$ if and only if there is an edge between nodes $i$ and $j$. For convenience, the graph is often referred to as $A$.

\begin{definition}[Adjacency spectral embedding]
    \label{def:ase}
    Suppose $\+A$ has the eigendecomposition $\+A = \sum_{i=1}^n \hat \lambda_i \hat u_i \hat u_i^\top$ with $|\hat \lambda_1| \geq \cdots \geq |\hat \lambda_n|$. The adjacency spectral embedding of the graph into $\R^r$, denoted $\hat X_1,\ldots,\hat X_n \in \R^r$, is given by the rows of the matrix
    \begin{equation*}
      \hat{\+X} =
      \begin{pmatrix}
        \hat X_1^\top \\
        \vdots \\
        \hat X_n^\top
      \end{pmatrix} 
      := \brackets{|\hat \lambda_1|^{1/2}\hat u_1 \: \cdots \: |\hat \lambda_r|^{1/2} \hat u_r}
    \end{equation*}
    obtained by stacking the scaled eigenvectors $|\hat \lambda_1|^{1/2}\hat u_1,  \ldots, |\hat \lambda_r|^{1/2} \hat u_r$ in columns.
\end{definition}

The geometry of the vectors obtained from the adjacency spectral embedding of $A$ can be interpreted via a generic low-rank random graph model known as the generalised random dot product graph \citep{rubin2022statistical}, in which edges occur independently of one another and edge  probabilities are modelled using the indefinite inner product $\inner{\cdot,\cdot}_{p,q}$, defined by
\begin{equation*}
  \label{eq:indefinite_inner_product}
  \inner{x,y}_{p,q} := \sum_{i=1}^p x_i y_i - \sum_{i=p+1}^{p+q} x_i y_i \equiv x^\top \Ipq y, 
\end{equation*}
for $x,y \in \R^{p+q}$, where $\Ipq$ is the diagonal matrix of $p$ ones followed by $q$ minus-ones.

 \begin{definition}[Generalised random dot product graph]
 \label{def:GRDPG}
     The graph $A$ is said to follow a \emph{generalized random dot product graph} model with latent positions $X_1,\ldots,X_n \in \R^r$ and signature $(p,q)$ if $\{a_{ij}\}_{i<j}$ are independent Bernoulli random variables with success probabilities
     \begin{equation*}
         p_{ij} := \inner{X_i, X_j}_{p,q} \in [0,1], \qquad 1 \leq i < j \leq n.
     \end{equation*}
 \end{definition}

 \begin{remark}[Identifiability]
     There are two distinct sources of non-identifiability in the latent positions of a generalized random dot product graph. 
     First, one can increase the dimension of the latent positions, for example by padding them with zeroes, without changing the distribution of $A$. We preclude such parametrisations by requiring that $r = \rank(P)$, where $P := (p_{ij})$, or alternatively that the second-moment matrix $ \Delta := n^{-1}\sum_{i=1}^n X_i X_i^\top$ has full rank.
     Second, replacing $\{X_i\}_{i=1}^n$ with $\{Q X_i\}_{i=1}^n$ for any matrix $Q$ in the indefinite orthogonal group $\O(p,q) = \smallcurlybrackets{M : M^\top \Ipq M = \Ipq}$ does not change $P$. If $P$ has distinct eigenvalues, the model can be made identifiable by requiring, for example, that $\Delta$ be diagonal \citep{agterberg2020two}, however we prefer to consider $\{Q X_i : Q \in \O(p,q)\}_{i=1}^n$ as an equivalence class of latent positions, and the theory we present reflects this.
 \end{remark}

 \begin{remark}[Signature]
     The random dot product graph model \citep{young2007random} is a special case of Definition~\ref{def:GRDPG} in which $q=0$. We remark that the edge probability matrix of a multipartite graph necessarily has negative eigenvalues, and therefore cannot be modelled using a random dot product graph model.
 \end{remark}

 The following theorem, which is a corollary of Corollary~3.1 of \citet{xie2021entrywise} (see also Corollary~4.1 of \citet{xie2024entrywise} and Theorem~1 of \citet{rubin2022statistical}), asserts that provided the latent positions $\{X_i\}_{i=1}^n$ are of the same order of magnitude, and the expected degrees of the graph grows at least logarithmically in $n$, then the adjacency spectral embedding of $A$, $\{\hat X_i\}_{i=1}^n$ provide a uniformly consistent estimate of the latent positions $\{X_i\}_{i=1}^n$. Here, the embedding dimension is equal to the rank $r = p+q$ which is considered known.

 % \begin{theorem}[Corollary of Corollary~3.1 of \citet{xie2021entrywise}]
\begin{theorem}[\citet{xie2021entrywise}]
 \label{thm:xie_consistency}
     Suppose $A$ follows a generalized random dot product graph with signature $(p,q)$ and latent positions $\{X_i\}_{i=1}^n$ satisfying $\smallnorm{X_i}_2 \eqc \rho^{1/2}$ for some $\rho \leqc 1$. Then, provided $\kappa(\Delta) := \sigma_1(\Delta) / \sigma_r(\Delta) \eqc 1$, $r \eqc 1$ and $n\rho \geqc \log n$, there exists an indefinite orthogonal matrix $Q \in \O(p,q)$ such that, with overwhelming probability,
     \begin{equation*}
         \max_{i\in[n]} \norm{\hat X_i - Q X_i}_2 \leqc \brackets{\frac{\log n}{n}}^{1/2}.
     \end{equation*}
 \end{theorem}

 For this reason, for large graphs, we can expect the geometry of $\smallcurlybrackets{\hat X_i}_{i=1}^n$ to approximately reflect that of $\smallcurlybrackets{X_i}_{i=1}^n$.

% \begin{remark}
%     When the optional Laplacian normalisation step of Remark~????????????????????????????????????????????????????????????\ref{remark:adj_rescaling} is used, the estimand in Theorem~????????????????????????????????????????????????????????????\ref{thm:xie_consistency} becomes {E(δi)+τ}−1/2Xi\{E(\delta_i) + \tau\}^{-1/2} X_i. Since this is simply a rescaling of XiX_i, the subspace geometry explored in the proceeding section applies.
% \end{remark}

\subsection{Spectral embedding of bipartite graphs}
\label{sec:bipartite_graphs}

A graph is said to be bipartite if its vertex set can be partitioned into two disjoint subsets as $V \equiv V_1 \cup V_2$ such that $a_{ij} = 0$ if $i,j \in V_1$ or $i,j \in V_2$. We refer to the partition membership of a node as its ``type'' and denote the type of node $i$ by $z_i \in \{1,2\}$. We assume without loss of generality that the nodes are indexed such that $z_1 \leq \cdots \leq z_n$ and write $n_k = |V_k|$. Then, the graph adjacency matrix has the form
\begin{equation*}
    A = \begin{pmatrix}
        0 & \*A \\ \*{A}^\top & 0
    \end{pmatrix},
\end{equation*}
where $\*A \in \{0,1\}^{n_1 \times n_2}$ is known as the graph biadjacency matrix whose rows correspond to $V_1$ and whose columns correspond to $V_2$. While an embedding of $A$ can be obtained using Definition~\ref{def:ase}, it is more common to approximately factorize $\*A$ using a truncated singular value decomposition, known as biadjacency spectral embedding.

\begin{definition}[Biadjacency spectral embedding]
  \label{def:bi_ase}
  Suppose $\*{A}$ has the singular value decomposition $\*{A} = \sum_{i=1}^n \hat s_i \hat u_i \hat v_i^\top$ with $\hat s_1 \geq \cdots \geq \hat s_n$. The biadjacency spectral embedding of $\*{A}$ into $\R^d$, denoted $\hat Y_1,\ldots, \hat Y_n$, is given by the rows of the matrices
  \begin{equation*}
    \hat{\+Y}^{(1)} \equiv
    \begin{pmatrix}
        \hat Y_1^\top \\
        \vdots \\
        \hat Y_{n_1}^\top
      \end{pmatrix} 
     := \brackets{\hat s_1^{1/2}\hat u_1 \: \cdots \: \hat s_d^{1/2} \hat u_d}, \qquad \hat{\+Y}^{(2)} \equiv 
     \begin{pmatrix}
        \hat Y_{n_1+1}^\top \\
        \vdots \\
        \hat Y_n^\top
      \end{pmatrix}
      := \brackets{\hat s_1^{1/2}\hat v_1 \: \cdots \: \hat s_d^{1/2}\hat v_d},
  \end{equation*}
  obtained by stacking the scaled left singular vectors $\hat s_1^{1/2} \hat u_1, \ldots, \hat s_d^{1/2} \hat u_d$, and scaled right singular vectors  $\hat s_1^{1/2} \hat v_1, \ldots, \hat s_d^{1/2} \hat v_d$, respectively, in columns.
\end{definition}

The matrix $A$ is known in the literature as the symmetric dilation of $\*A$, and the eigenvalues and vectors of $A$, and the singular values and vectors of $\*A$ have the following geometric relationship:
% \footnote{It was used in Jordan's construction of the singular value decomposition in 1874 \citep{stewart1990matrix}.}

\begin{proposition}[Theorem 7.3.3 of \citet{stewart1990matrix}]
Suppose $s$ is a singular value of $\*A$ and $u, v$ are corresponding left and right singular vectors, then $\pm s$ are eigenvalues of $A$ and
\begin{equation*}
    \frac{1}{\sqrt{2}}\begin{pmatrix}
        u \\ \pm v
    \end{pmatrix}
\end{equation*} are corresponding eigenvectors.
\end{proposition}

\begin{figure}%[tbhp]
  \centering
  \begin{overpic}[height=16em]{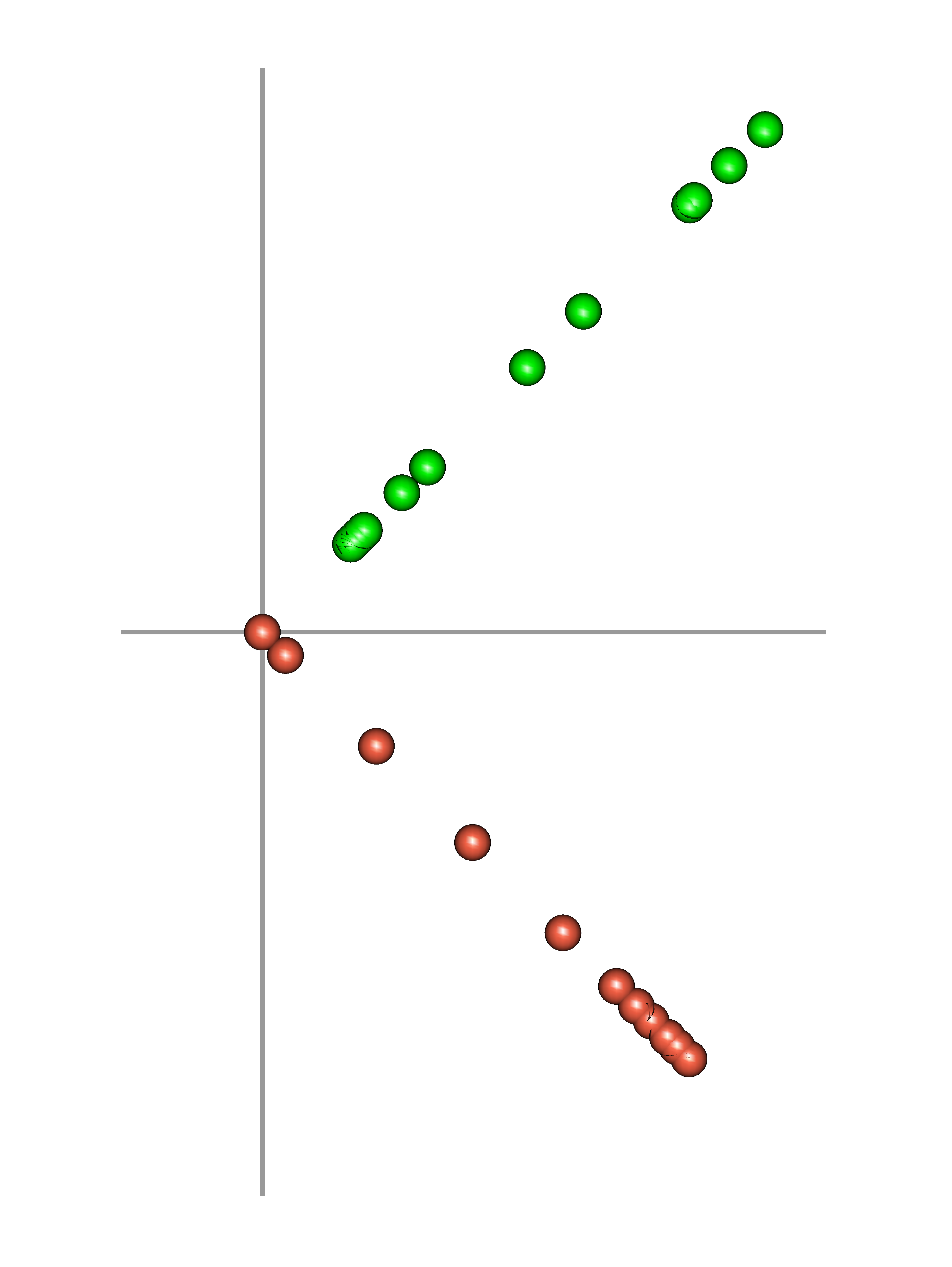}
    \put (-9,53) {\footnotesize$\hat{X}_1,\ldots,\hat X_n$}
    \put (16,45) {\tiny \textcolor{gray}{$0$}}
   \end{overpic}
   \hspace{1em}
  \begin{overpic}[height=16em]{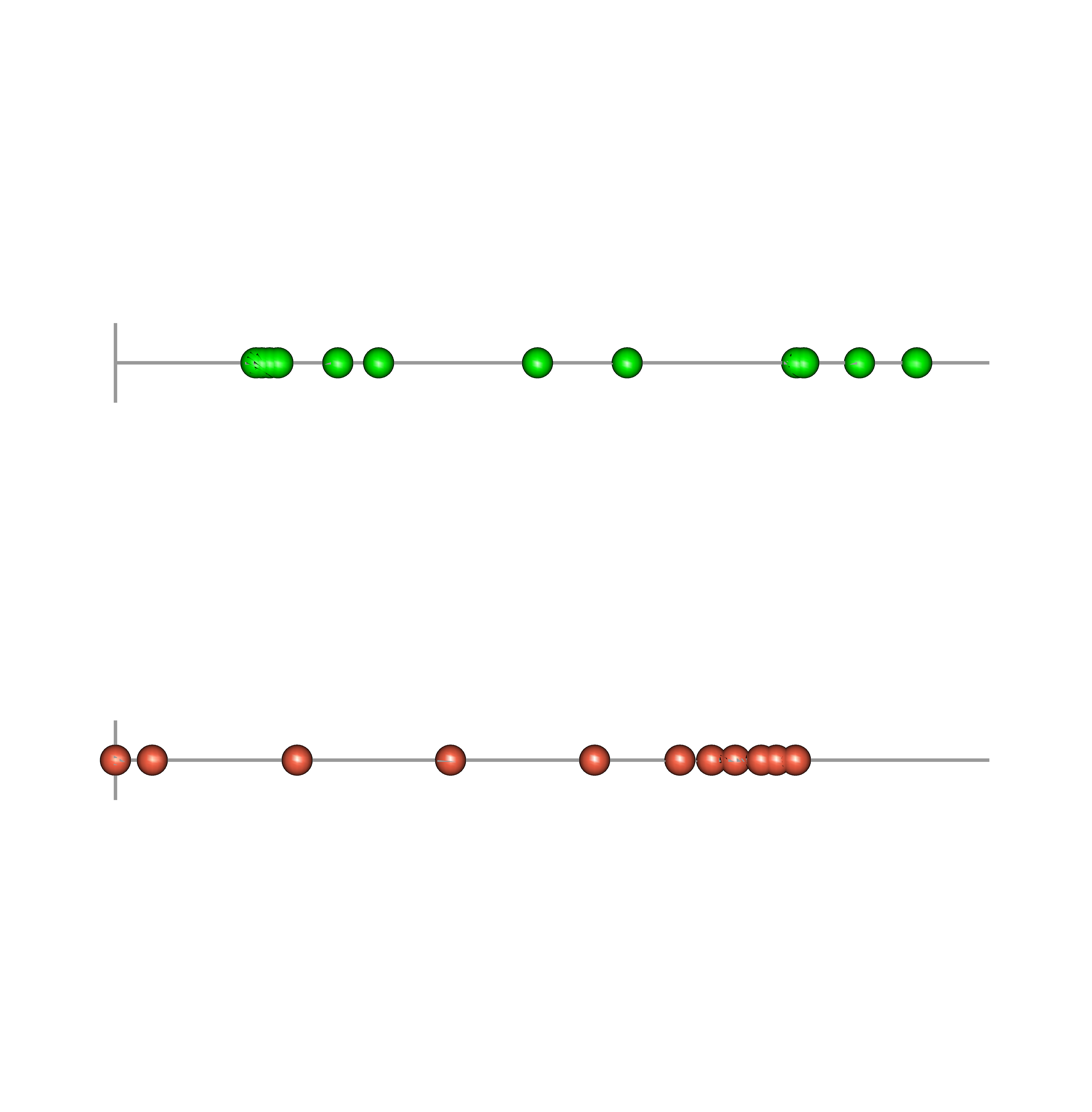}
    \put (10,75) {\footnotesize$\hat{Y}_1,\ldots,Y_{n_1}$}
    \put (10,39) {\footnotesize$\hat{Y}_{n_1 + 1},\ldots,\hat Y_n$}
    \put (9.15,59) {\tiny \textcolor{gray}{$0$}}
    \put (9.15,23.5) {\tiny \textcolor{gray}{$0$}}
   \end{overpic}
  \vspace*{-1em}
  \caption{The latent geometry of a bipartite graph. The left panel shows the two-dimensional spectral embedding, $\hat{X}_1,\ldots,\hat X_n$, of a bipartite graph generated from the random graph model \eqref{eq:chung_lu} where $z_i \in \{1,2\}$, coloured by type. The right panel shows the corresponding one-dimensional biadjacency spectral embedding, $\hat{Y}_1,\ldots,\hat{Y}_n$ of the graph.}
  \label{fig:bi_chung_lu}
\end{figure}

The proof of this statement is a simple computation and it implies the following geometric relationship between $\smallcurlybrackets{\hat X_i}_{i=1}^n$ and $\smallcurlybrackets{\hat Y_i}_{i=1}^n$:
\begin{lemma}
\label{lemma:bipartite_generalization}
    Let $\smallcurlybrackets{\hat X_i}_{i=1}^n$ be the adjacency spectral embedding of $A$ into $\R^{2d}$ (Definition~\ref{def:ase}) and let $\smallcurlybrackets{\hat Y_i}_{i=1}^n$ be the biadjacency spectral embedding of $\*A$ into $\R^d$ (Definition~\ref{def:bi_ase}). Then, for compatibly chosen spectral decompositions,
    \begin{equation*}
        \hat Y_i =
            \frac{1}{\sqrt{2}}\begin{pmatrix}
                \hat X_i \\ \hat X_i
            \end{pmatrix} \quad \text{ for } i \in V_1; \qquad
        \hat Y_i =\frac{1}{\sqrt{2}}\begin{pmatrix}
                \hat X_i \\ -\hat X_i 
            \end{pmatrix} \quad \text{ for } i \in V_2.
    \end{equation*}
\end{lemma}
The adjacency spectral embedding is exactly an embedding of the two sides of the biadjacency spectral embedding into orthogonal $d$-dimensional subspaces of $\R^{2d}$. Figure~\ref{fig:bi_chung_lu} illustrates Lemma~\ref{lemma:bipartite_generalization} with a toy example.

\section{The latent geometry of multipartite networks}
% A graph is said to be multipartite if its node set can be partitioned into $K$ disjoint sets $V = V_1 \cup \cdots \cup V_K$, each corresponding to a type of node, such that $a_{ij} = 0$ if $i$ and $j$ are of the same type. As before, we write $n_k = |V_k|$, we denote a node's type by $z_i \in [K]$ and assume without loss of generality that the nodes are indexed such that $z_1 \leq \cdots \leq z_n$. For the rest of the paper, we assume $A$ is multipartite and that the node partitioning is known.

A graph is said to be multipartite if its node set can be partitioned into $K$ disjoint sets $V = V_1 \cup \cdots \cup V_K$ such that $a_{ij} = 0$ if $i, j \in V_k$ for some $k \in [K]$. As before, we refer to the partition membership of a node as its ``type'', we denote the type of node $i$ by $z_i \in [K]$ and write $n_k = |V_k|$. For the rest of the paper, we assume $A$ is multipartite and that the node partitioning is known.

\subsection{A tripartite example}
To motivate our discussion of multipartite random graphs, we consider a simple tripartite inhomogeneous random graph, in which each node $i \in [n]$ is assigned a scalar-valued weight, $w_i \in (0,1]$, and edge probabilities are given by
\begin{equation}
    p_{ij} = \begin{cases}
    w_i w_j & \text{ if } z_i \neq z_j, \\
    0 & \text{ if } z_i = z_j.
    \end{cases}
    \label{eq:chung_lu}
\end{equation}
This model is a multipartite generalisation of the Chung-Lu model \citep{aiello2001random,chung2002average,chung2002connected}.
The edge probability matrix $P$ has rank three, and is equivalently described using a generalised random dot product graph with signature $(1,2)$, and latent positions $X_i = w_i \alpha_{z_i} (1, \cos \theta_{z_i}, \sin \theta_{z_i})^\top$ with distinct angles $\theta_1,\theta_2,\theta_3 \in [0,2\pi)$ and compatibly defined $\alpha_1,\alpha_2,\alpha_3\in \R$.

The latent positions lie on three one-dimensional subspaces, corresponding to the three node types. These subspaces are necessarily \emph{totally isotropic} with respect to the indefinite inner product $\langle \cdot, \cdot \rangle_{p,q}$, meaning that the indefinite inner product of any two points on a subspace is zero.

\begin{figure}
% The arguments in the next line are {height}{optional width}{used only by OUP for typesetting}[filename, in directory art]
% \figurebox{20pc}{25pc}{}[XP.png]
\centering
\includegraphics[width=0.45\linewidth, trim={28pt 48pt 28pt 32pt}, clip]{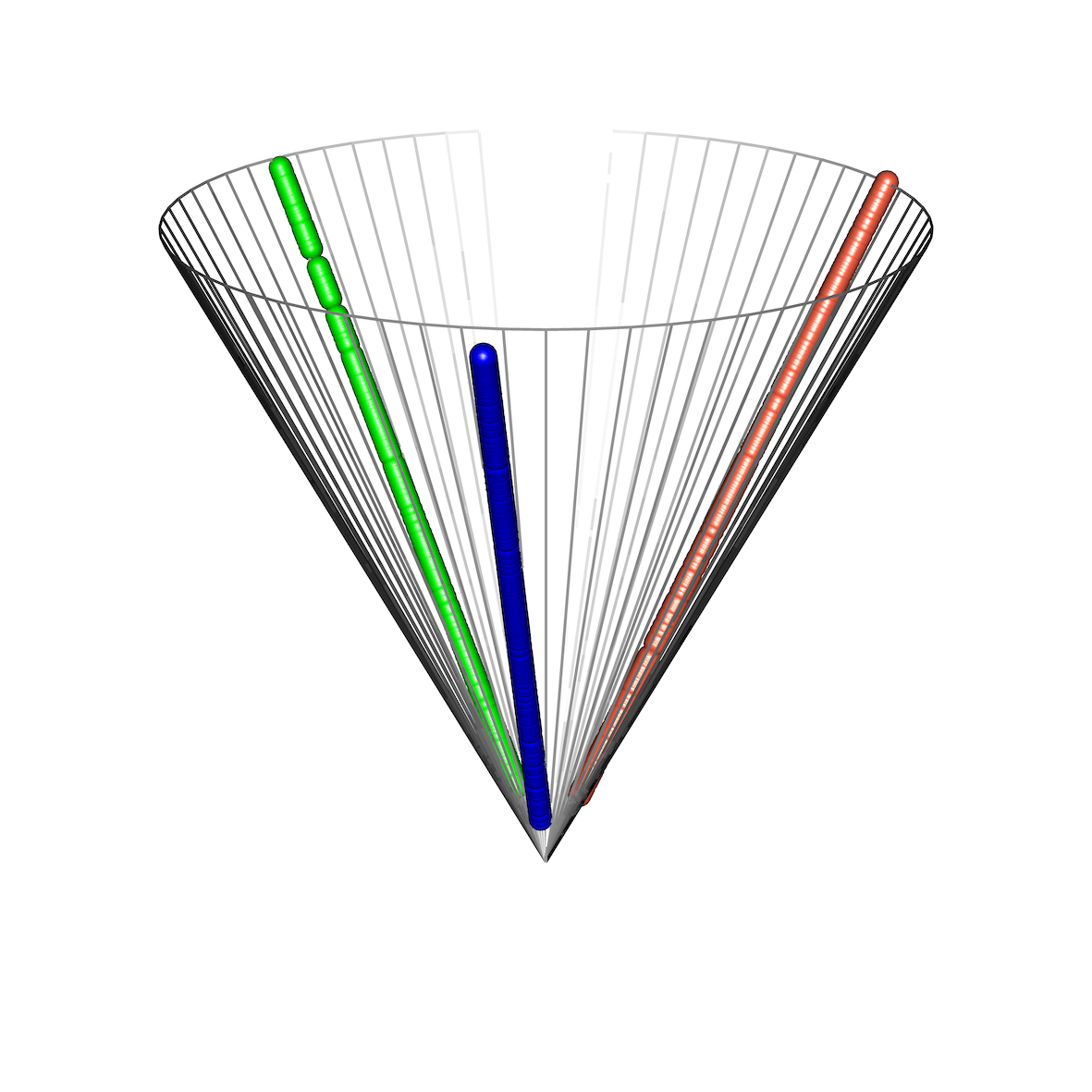}
\includegraphics[width=0.45\linewidth, trim={28pt 48pt 28pt 32pt}, clip]{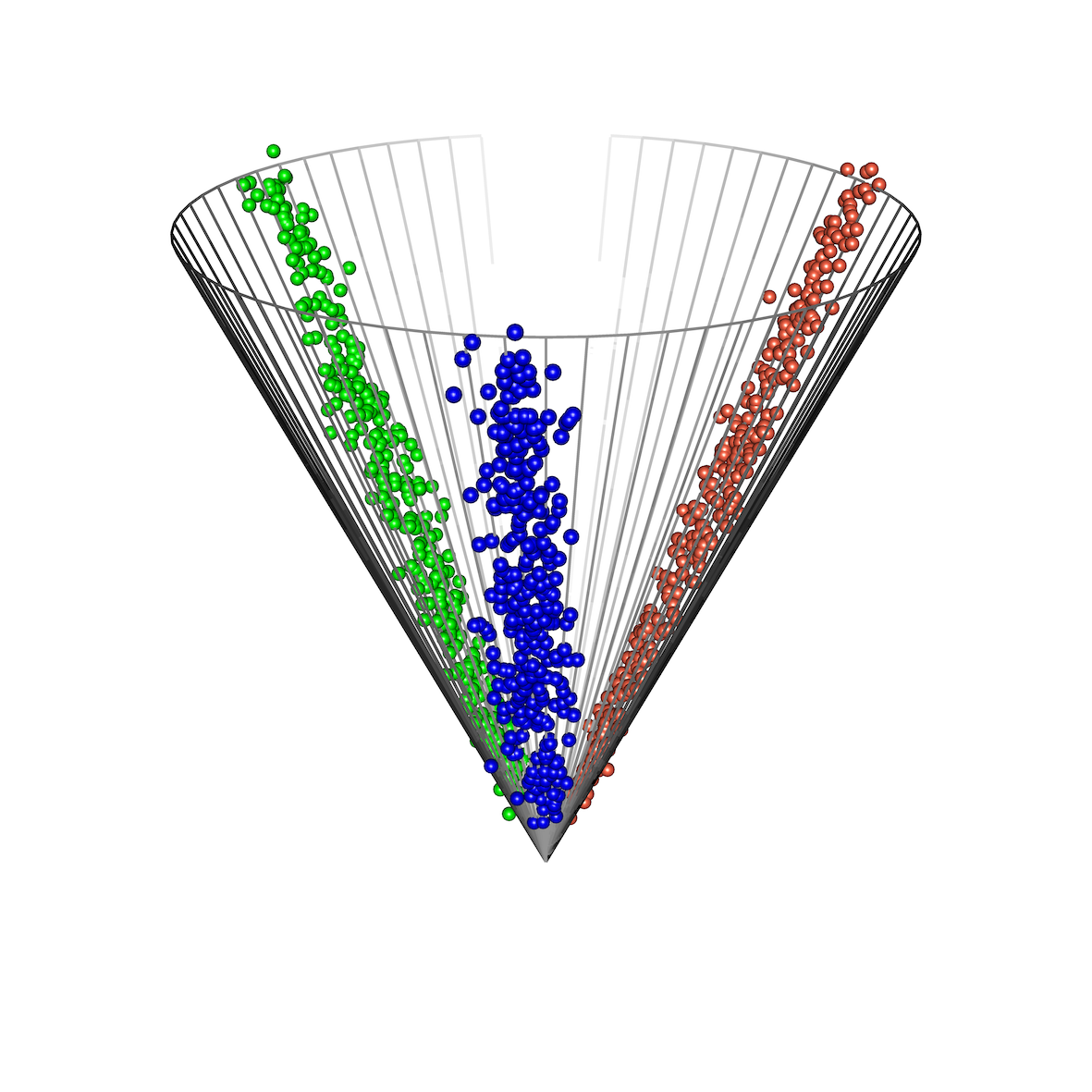}
% note that files may not be rotated
\caption{The latent geometry of a tripartite graph. The left panel shows three-dimensional latent positions $X_1,\ldots,X_n$ corresponding to the tripartite random graph described in \eqref{eq:chung_lu}, coloured by type. }
% The cone represents all the totally isotropic subspaces in $\R^3$ with signature $(1,2)$. The right panel shows the spectral embedding of a simulated realization of such a graph.}
\label{fig:chung_lu}
\end{figure}

The left panel of Figure~\ref{fig:chung_lu} shows a configuration of these three-dimensional latent positions with 300 nodes of each type and weights drawn uniformly on the interval $[0.1,1]$. The cone represents all totally isotropic subspaces in $\R^3$ with signature $(1,2)$. The right panel shows the adjacency spectral embedding of a simulated realization of the graph.

\subsection{Subspace geometry of multipartite graphs}
The subspace geometry observed in the previous example is a special case of a more general phenomenon.

\begin{lemma}
\label{lemma:subspacec_dimension}
    The latent positions $\{X_i\}_{i \in V_k}$ corresponding to nodes of a single type $k \in [K]$ lie on a subspace which is totally isotropic with respect to the indefinite inner product $\inner{\cdot,\cdot}_{p,q}$ and has dimension no greater that $\min\{p,q\}$.
\end{lemma}

Lemma~\ref{lemma:subspacec_dimension} follows as a corollary of Witt's theorem of quadratic forms \citep{lam2005introduction}. An elementary proof is given in the appendix, along with all other proofs in this paper.
We denote the dimension of the totally isotropic subspace supporting $\{X_i\}_{i \in V_k}$ by $r_k$, and we refer to $r$ as the \emph{ambient dimension} and $r_1,\ldots,r_K$ as the \emph{intrinsic dimensions} of the graph.

\section{Spectral embedding of multipartite graphs}
% To summarise what we've learned so far: Lemma~\ref{lemma:xy_equiv} states that the adjacency spectral embedding of a bipartite graph into $2d$ dimensions lies exactly of the union of two $d$-dimensional subspace, and a combination of Lemma~\ref{lemma:subspacec_dimension} and Theorem~\ref{thm:xie_consistency} show that the 

% Under a multipartite, low-rank random graph model, spectral embedding gives a representation of the nodes in their ambient rather than intrinsic dimension. To obtain intrinsic-dimensional representations, we propose a subsequent, group-specific, subspace projection step. In statistical parlance, we perform group-specific, uncentered, principal component analysis.

A combination of Lemma~\ref{lemma:subspacec_dimension} and Theorem~\ref{thm:xie_consistency} says that the adjacency spectral embedding of a low-rank multipartite random graph lies ``close'' to the union of $K$ lower-dimensional subspaces. This motivates a subsequent step to the standard spectral embedding algorithm of estimating these subspaces and projecting onto them, to obtain vector representations of the nodes of each type in their intrinsic, rather than ambient, dimension. We propose to estimate these subspaces using type-specific, uncentered principal component analysis.
\begin{definition}[Multipartite adjacency spectral embedding]
    \label{def:mpspectral}
    Given positive integers $r_1,\ldots,r_K\leq r$, let $\hat{X}_1,\ldots, \hat X_n$ be the adjacency spectral embedding of $A$ into $\R^r$ (see Definition~\ref{def:ase}). Then the \emph{multipartite adjacency spectral embedding} of node $i \in V_k$ into $\R^{r_k}$ is \[\hat{Y}_i = \hat{\Xi}_k^\top\hat{X}_i,\]
    where the columns of $\hat{\Xi}_k$ are the $r_k$ eigenvectors of $\hat \Sigma_k := n_k^{-1}\sum_{i \in V_k} \hat X_i \hat X_i^\top$ with the largest eigenvalues.
    % is obtained as:  % with  $\sum_k d_k \geq D$, 
    % \begin{enumerate}
    % \item \emph{Spectral embedding.} Let $\hat{X}_1,\ldots, \hat X_n$ be the adjacency
    % or Laplacian
    % spectral embedding of $A$ into $\R^r$ (see Definition~\ref{def:ase}),
    % divided into group-specific point clouds,
    % \[\hat{X}^{(k)} := (\hat X_{i})^{\top}_{i \in V_k}.\]
    % where $i_1, \ldots, i_{n_{k}}$ are the ordered indices of the nodes for which $z_i = k$, for $k \in \{1, \ldots, K\}$.
    % \item \emph{Dimension reduction.} The \emph{multipartite spectral embedding} of node $i \in V_k$ into $\R^{d_k}$ is
    % \[\hat{Y}_i = \brackets{\hat{\Xi}^{(k)}}^\top\hat{X}_i,\]
    % where the columns of $\hat{\Xi}^{(k)}$ are $d_k$ eigenvectors of $\hat \Sigma^{(k)} := \sum_{i \in V_k} \hat X_i \hat X_i^\top$ with the largest eigenvalues.
    % \end{enumerate}
\end{definition}

As indicated by the notation, Definiton~\ref{def:mpspectral} contains biadjacency spectral embedding (Definition~\ref{def:bi_ase}) as a special case when $d_1 = d_2 =: d$ and $r = 2d$, which can be proved using the observations of Section~\ref{sec:bipartite_graphs}.

In modern applications where node degrees are highly heterogeneous, the normalisation step in the next remark is often considered before the spectral decomposition \citep{chaudhuri2012spectral, amini2013pseudo, qin2013regularized}.

\begin{remark}[Optional Laplacian normalisation]
\label{remark:adj_rescaling}
    Let $d_i = \sum_{j=1}^n a_{ij}$ denote the degree of node $i \in [n]$, set a regularisation parameter $\tau \geq 0$ and define the matrix $D_\tau = \diag(d_1 + \tau, \ldots, d_n + \tau)$. Define the normalised Laplacian matrix $L_\tau = D_\tau^{-1/2}A D_\tau^{-1/2}$, and replace $A$ in Definition~\ref{def:ase} with $L_\tau$.
\end{remark}

In the literature, $L_\tau$ is known as the regularised Laplacian matrix, and is equal to the standard normalised Laplacian matrix when $\tau = 0$, and properly scaled, converges to the adjacency matrix when $\tau \to \infty$. For a properly chosen $\tau$, it has been shown to improve the statistical performance of the spectral embedding for sparse graphs \citep{le2017concentration, zhang2018understanding}. When the Laplacian normalisation step is applied, the population quantity in Theorem~\ref{thm:xie_consistency} becomes $\{E(d_i) + \tau\}^{-1/2} X_i$, a rescaling of $X_i$, and therefore the subspace geometry explored in the preceding section still applies. For simplicity of analysis, we do not consider this step in the theory of this paper, however we do in the real data example of Section~\ref{sec:real_data}.

\subsection{Estimation theory}

To facilitate statistical analysis of point clouds obtained by multipartite spectral embedding, we put down a random graph model parametrised to make the true, instrinsic-dimensional representations of the nodes explicit. 

\begin{definition}[Multipartite bilinear random graph model]
    Let  $\Lambda_{k,\ell}$ be fixed, $r_k \times r_\ell$ matrices satisfying $\Lambda_{k,\ell} = \Lambda_{\ell,k}^\top$ for all $k \neq \ell$ and $\Lambda_{k,\ell} = 0$ for all $k = \ell$. Let $\{Y_i\}_{i\in V_k}$ be vectors in $\R^{r_k}$ satisfying $Y_i^\top \Lambda_{z_i,z_j} Y_j \in [0,1]$. The graph $A$ follows a \emph{multipartite bilinear random graph model} with link matrices $\{\Lambda_{k,\ell}\}_{k < \ell}$ if $\smallcurlybrackets{a_{ij}}_{i<j}$ are independent Bernoulli random variables with success probabilities
    \begin{equation*}
        p_{ij} = Y_i^\top \Lambda_{z_i,z_j} Y_j, \qquad 1 \leq i < j \leq n.
    \end{equation*}
\end{definition}

While the generalised random dot product graph model requires that edge probabilities be modelled by a common function of the latent positions (i.e. the indefinite inner product), the multipartite bilinear random graph model allows edges between nodes in different pairs of blocks to be modelled using \emph{different} functions. This additional flexibility is what allows nodes to be parametrized in their intrinsic dimension.

\begin{remark}[Identifiability]
    There are two distinct sources of non-indentifiability in the the latent positions and link matrices of a multipartite inner-product model. First, as with the generalised random dot product graph, one can increase the dimension of the latent positions, for example by padding them with zeros without changing $P$. We preclude this by requiring that the matrices $\Lambda_k := \brackets{\Lambda_{k1} \: \cdots \: \Lambda_{kK}}$ and $\Sigma_k := n_k^{-1}\sum_{i\in V_k} Y_i Y_i^\top$ have full rank. Second, replacing $\{Y_i\}_{i\in V_k}$ with $\{\+G_k Y_i\}_{i\in V_k}$ and $\smallcurlybrackets{\+\Lambda_{k\ell}}_{k,\ell=1}^K$ with $\smallcurlybrackets{(\+G_k^\top)^{-1}\+\Lambda_{k\ell}\+G_\ell^{-1}}_{k,\ell=1}^K$, where $\{\+G_k\}_{k=1}^K$ are invertible, linear transformations, does not change the distribution of $\+A$. Therefore the latent positions are only identified up to type-wise invertible transformations, a fact reflected in the consistency result to come.
    
    % We impose the condition that for all k∈[K]k\in[K], the matrices Λk:=\bracketsΛk1⋯ΛkK\Lambda_k := \brackets{\Lambda_{k1} \: \cdots \: \Lambda_{kK}} and Δk:=∑i∈VkYiY⊤i\Delta_k := \sum_{i\in V_k} Y_i Y_i^\top have full rank, which ensures that the link matrices and latent positions are parameterised in the smallest dimension possible. Replacing {Yi}i∈Vk\{Y_i\}_{i\in V_k} with {GkYi}i∈Vk\{G_k Y_i\}_{i\in V_k} and \smallcurlybracketsΛkℓ\smallcurlybrackets{\Lambda_{k\ell}} with \smallcurlybrackets(G⊤k)−1ΛkℓG−1ℓ\smallcurlybrackets{(G_k^\top)^{-1}\Lambda_{k\ell}G_\ell^{-1}} for Gk∈\GL(dk),Gℓ∈\GL(dℓ),where,where\GL(dk)denotesthegroupofinvertiblelineartransformations,foralldenotesthegroupofinvertiblelineartransformations,forallk,ℓ∈[K],doesnotchangethedistributionof,doesnotchangethedistributionofAG_k \in \GL(d_k), G_\ell \in \GL(d_\ell),where, where \GL(d_k)denotesthegroupofinvertiblelineartransformations,forall denotes the group of invertible linear transformations, for all k,\ell \in [K],doesnotchangethedistributionof, does not change the distribution of A.
\end{remark}

% -- define \Lambda !

The following theorem asserts that, provided the latent positions $\{Y_i\}_{i=1}^n$ are of the same order of magnitude, the expected degrees of the graph grow at least logarithmically in $n$, and the model is well-conditioned, then the multipartite adjacency spectral embedding of $\+A$, $\{\hat Y_i\}_{i=1}^n$, with ambient embedding dimension $r := \rank(\+\Lambda)$ and intrinsic embedding dimensions equal to the dimension of the corresponding latent positions,  provides a uniformly consistent estimate of the latent positions $\{Y_i\}_{i=1}^n$. By ``well-conditioned'', we mean that the matrices $\+\Sigma_k, \+\Lambda_k$ and $\+\Lambda$, the latter defined as the block-structured matrix whose $k\ell$th block is $\Lambda_{k\ell}$, have (reduced) condition numbers
\begin{equation*}
    \kappa(\+\Sigma_k) = \frac{\sigma_1(\+\Sigma_k)}{\sigma_{r_k}(\+\Sigma_k)}, \qquad \kappa(\+\Lambda_k) = \frac{\sigma_1(\+\Lambda_k)}{\sigma_{r_k}(\+\Lambda_k)}, \qquad \kappa(\+\Lambda) = \frac{\sigma_1(\+\Lambda)}{\sigma_{r}(\+\Lambda)},
\end{equation*}
respectively, which are of constant order for all $k\in[K]$. Here, we assume that the ambient and intrinsic dimensions are  known.

\begin{theorem}
\label{thm:mp_consistency}
    Suppose $\+A$ follows a multipartite bilinear random graph model with link matrices $\{ \+{\Lambda}_{k,\ell}\}_{k,\ell=1}^{K}$ and latent positions $\{Y_i\}_{i=1}^n$ satisfying $\smallnorm{Y_i}_2 \eqc \rho^{1/2}$ for some $\rho \leqc 1$. Then, providing 
    $$\kappa(\+\Sigma_k) \eqc \kappa(\+\Lambda_k) \eqc \kappa(\+\Lambda) \eqc 1,\qquad n_k \eqc n, \qquad r_k \eqc 1,$$
    for all $k \in [K]$, and $n\rho \geqc \log n$, there exist invertible matrices $\{\+G_{k}\}_{k=1}^K$ such that with overwhelming probability,
    \begin{equation*}
        \max_{i \in [n]} \norm{ \hat Y_i - \+G_{z_i} Y_i}_2 \leqc \brackets{\frac{\log n}{n}}^{1/2}.
    \end{equation*}
\end{theorem}

We remark here that the ``low-rank'' and ``well-conditioned'' assumptions can likely be relaxed, however this would come at the cost of significantly more complicated and less easily interpretable theoretical statements. We refer the interested reader to the dedicated literature on entrywise eigenvectors bounds under signal-plus-noise matrix models for details of how these assumptions might be relaxed 
\citep{cape2019signal, cape2019two, lei2019unified, abbe2020entrywise, xie2024entrywise}.

% derive in appendix

\subsection{Selecting the embedding dimension}
\label{sec:dim_select}

The estimation theory in this paper assumes that the embedding dimensions, both ambient and intrinsic, are known, and correspond to the population ranks of the adjacency matrix and the ambient latent positions corresponding to each type. This represents an ``unrealistic ideal'' in two respects. First, in practice these dimensions need to be selected by the practitioner using the data. Second, the finite rank assumption on $P$ might not be expected to hold exactly. As a result, we prefer to view practical dimension selection as a bias/variance trade-off rather than an estimation problem. Indeed, even if we knew the ambient and intrinsic dimensions, in finite samples they might not be the best to choose. We refer the reader to \citet{chen2021estimating,priebe2019two} and \citet{whiteley2022discovering} for pragmatic discussions around this topic. Several rank selection methods are available in the literature \citep{zhu2006automatic,luo2016combining,chen2021estimating}. We use the elbow method of Zhu and Ghodsi \citep{zhu2006automatic} in our real data example, but leave the choice open in general.
% referring to a generic rank selection function r(⋅)r(\cdot) in Algorithm~????????????????????????????????????????????????????????????????????????????????????????????????????????????????????????????????????????????????\ref{alg:clustering}.
Note that Lemma~\ref{lemma:subspacec_dimension} provides the maximal value of the intrinsic dimensions given the ambient dimension. We have found that reasonable rank selection procedures rarely break this inequality in practice (e.g. see Figure~\ref{fig:scree_plots}). 

% \newpage

\section{Spectral clustering}
A common use of spectral embedding is to uncover community structure in a network. This is achieved via a subsequent clustering procedure such as the $k$-means algorithm. Algorithm~\ref{alg:clustering} describes a spectral clustering algorithm for multipartite networks. It takes as input the embedding dimensions and number of communities which in practice must be estimated from the data.

\begin{algorithm}[!t]
%\label{alg:clustering}
 \caption{Multipartite spectral clustering}\label{alg:clustering}
 \begin{flushleft}
  \textbf{Input:} adjacency matrix $A$, node partition $V = V_1 \cup \cdots \cup V_K$, embedding dimensions $r_1,\ldots,r_K, r$, number of communities in each block $\kappa_1,\ldots,\kappa_K$.
 \end{flushleft}
\begin{algorithmic}[1]
  % \Statex \textbf{input} Multipartite graph A∈0,1n×nA \in {0,1}^{n \times n}, z1,…,zn∈{1,…,K}z_1, \ldots, z_n \in \{1,\ldots,K\}
  % \STATE Input multipartite graph A∈{0,1}n×nA \in \{0,1\}^{n \times n}, partition memberships z1,…,zn∈{1,…,K}z_1, \ldots, z_n \in \{1,\ldots,K\}
  \STATE Compute $\{\hat Y_i \}_{i\in V_k} \in \R^{d_k}$, $k\in[K]$, the multipartite spectral embedding of $A$ with ambient dimension $r$, and intrinsic dimensions $r_1,\ldots,r_K$ (Definition~\ref{def:mpspectral}).
  \STATE \emph{(optional)} For each $i \in [n]$, set $\hat{Y}_i = \hat{Y}_i/\|\hat{Y}_i\|$. 
  \STATE For each $k \in [K]$, apply $k$-means to $\{\hat Y_i\}_{i\in V_k}$ with $\kappa_k$ clusters.
  % \STATE Output community memberships ˆτ1,…,ˆτn\hat \tau_1, \ldots, \hat \tau_n. %Output community memberships ˆτi={∑k−1l=1^dl+ˆτi\hat{\tau}_i = \{\sum_{l=1}^{k-1} \hat{d_l} + \hat{\tau}_i, where zi=kz_i = k  
  %\Statex \Return dd-dimensional points ˆZ1,…,ˆZn\hat Z_1, \ldots, \hat Z_n
\end{algorithmic}
 \begin{flushleft}
  \textbf{Output:} community partition $\hat C_1 \cup \cdots \cup \hat C_\kappa = V$.
 \end{flushleft}
 \end{algorithm}

% \subsection{Consistency of multipartite spectral clustering}

The stochastic block model \citep{holland1983stochastic} and degree-corrected stochastic block model \citep{karrer2011stochastic} are models for community structured networks and are ubiquitous in the community detection literature. We formally define these models in the specific setting of multipartite graphs.

\begin{definition}[Multipartite degree-corrected stochastic block model]
    Suppose the vertex set $V = V_1 \cup \cdots \cup V_K$ of a multipartite graph $A$ is further sub-partitioned into $S$ disjoint blocks $V = C_1 \cup \cdots \cup C_\kappa$. Let $B \in [0,1]^{\kappa\times \kappa}$ be a matrix satisfying $b_{uv} = 0$ if $C_u, C_v \subseteq V_k$ for some $k \in [K]$, and let $w_1,\ldots,w_n \in [0,1]$ be a set of weights. The graph $A$ follows a \emph{multipartite degree-corrected stochastic block model} if $\smallcurlybrackets{a_{ij}}_{i<j}$ are independent Bernoulli random variables with success probabilities
    \begin{equation*}
        p_{ij} = w_i w_j b_{uv}, \qquad i \in C_u, j \in C_v.
    \end{equation*}
\end{definition}
If $w_1,\ldots,w_n = 1$, then the model is simply referred to as the multipartite stochastic block model. In the following, we let $m_k$ denote the number of communities of type $k$.

% All multipartite degree-corrected stochastic block models can be parametrised as a multipartite inner product graph model. One such parametrisation is to set Λk,ℓ\Lambda_{k,\ell} to the mk×mℓm_k \times m_\ell submatrix of BB corresponding to communities in blocks kk and ℓ\ell, and to set YiY_i to be the indicator vector with wiw_i in the position corresponding to its community.

% (xx) Comment about perfect clustering.

Any multipartite degree-corrected stochastic block models can be parametrised as a multipartite bilinear random graph model. When $\+B$ has full rank, one such parametrisation is to set $\+{\Lambda}_{k,\ell}$ to the $m_k \times m_\ell$ submatrix of $\+B$ corresponding to communities of type $k$ and $\ell$, and to set $Y_i$ to be the indicator vector with $w_i$ in the position corresponding to its community. In addition, in this case, the intrinsic dimensions are equal to the number of sub-communities of each type, and therefore intrinsic dimension estimates may also serve as estimates of these quantities.

A corollary of the uniform consistency result of Theorem~\ref{thm:mp_consistency} is that, supposing the graph follows a multipartite stochastic block model (and step 2 of Algorithm~\ref{alg:clustering} is not implemented) or a multipartite degree-corrected stochastic block model (and step 2 of Algorithm~\ref{alg:clustering} is implemented), then asymptotically almost surely, Algorithm~1 will perfectly estimate the community memberships, assuming the conditions of the theorem hold and that the number of communities of each type is known.
This may be proved using analogous arguments to those employed in \citet{lyzinski2014perfect}.

\begin{figure}[t]
\centering
\includegraphics[width=0.7\textwidth]{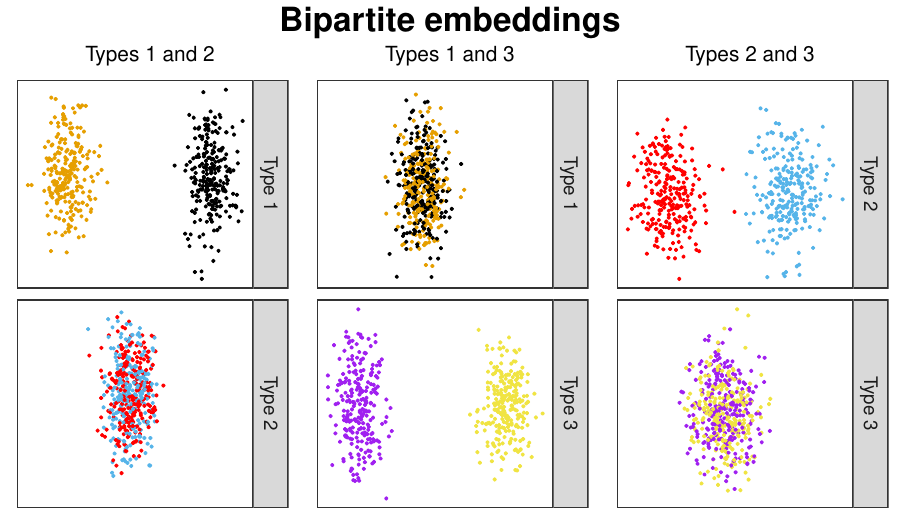}
    \includegraphics[width=0.7\textwidth]{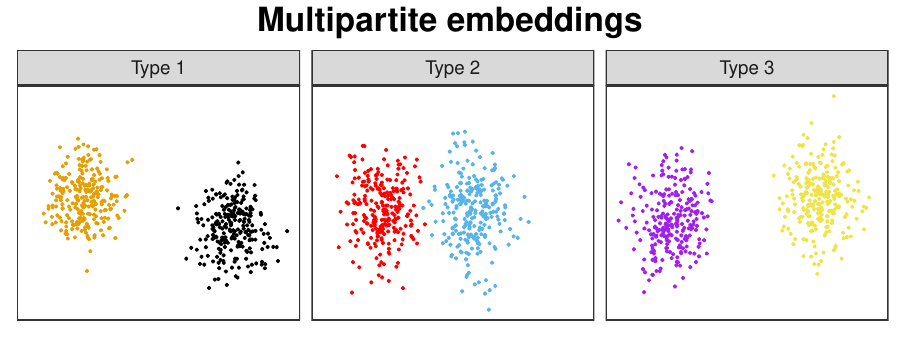}
    \caption{For a graph simulated from a multipartite stochastic block model with inter-community probability matrix of the form \eqref{eq:B_matrix}, the top panel shows the biadjacency spectral embeddings (Definition~\ref{def:bi_ase}) of the subgraphs corresponding to every pair of node types. For each pair of node types, two of the four relevant communities cannot be distinguished. The bottom panel shows the multipartite spectral embedding (Definition~\ref{def:mpspectral}) of the full tripartite graph, revealing all six communities.}
    \label{fig:obscured_sbm}
\end{figure}

\subsection{Obscured communities}

\begin{figure}[p]
  \centering
  \includegraphics[width=.7\linewidth]{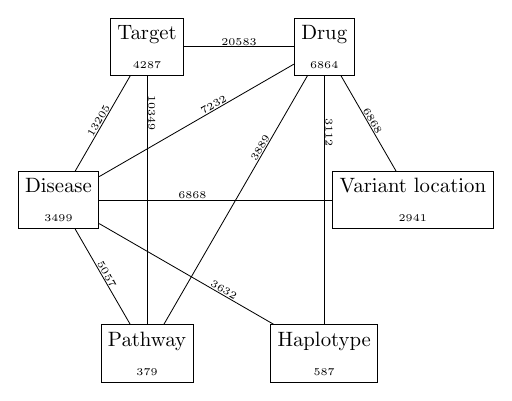}
  \caption{Schematic of the biomedical multipartite network analysed in Section~\ref{sec:real_data}. The number of nodes of each type and the number of edges between nodes of each type (zero if unspecified) are indicated.}
  \label{fig:biomedical_network_diagram}
\end{figure}

\begin{figure*}[p]
  \centering
  
  \begin{tabular}{c c}
     % \begin{subfigure}{0.29\textwidth}
      \centering
      % \hspace{4.7em}\small{\textbf{Ambient dimension selection}}\par\smallskip
      \includegraphics[height=12em]{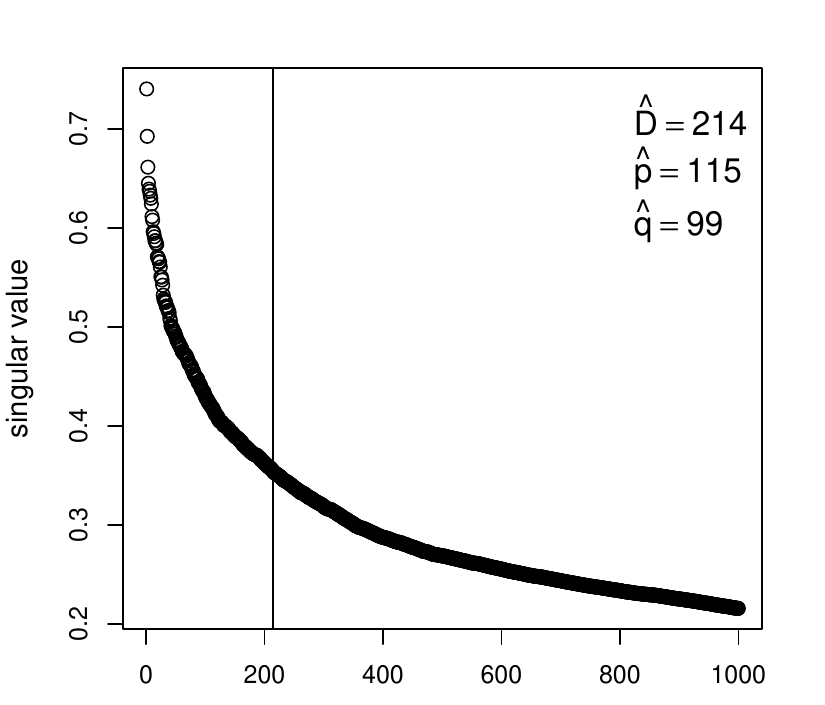}
  % \end{subfigure}
       &
  % \begin{subfigure}{0.69\textwidth}
      \centering
      % \hspace{2.9em}\small{\textbf{Intrinsic dimension selection}}\par\smallskip
      \includegraphics[height=12em]{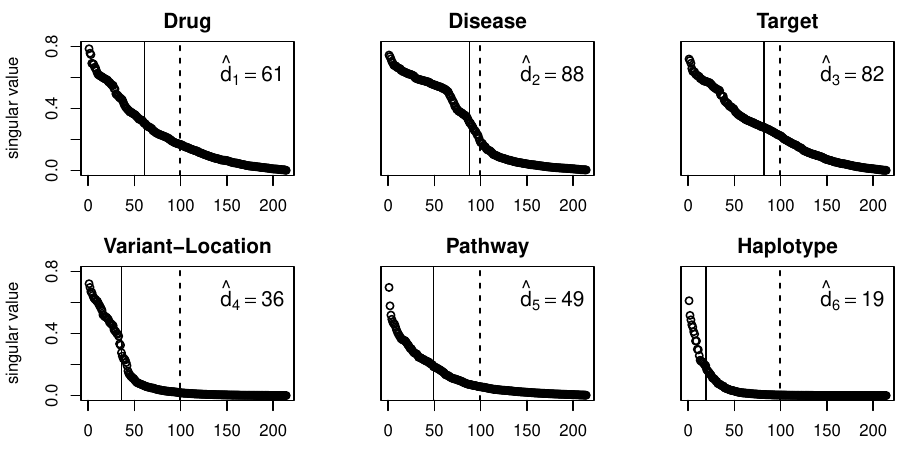}
  % \end{subfigure}
  \end{tabular}

  \caption{Dimension selection for the biomedical multipartite network. The left panel shows the scree plot of the regularized Laplacian matrix and the right panels the scree plots of the ambient embeddings corresponding to each type. The dimension selected --- and thus the number of clusters --- is shown as a solid line and $\min\{\hat p, \hat q\} = 99$ is shown as a dashed line.}
  \label{fig:scree_plots}
\end{figure*} 

The following is an example of a multipartite stochastic block model for which multipartite spectral clustering reveals the full community structure of a graph, but spectral clustering using any individual bipartite subgraph fails to reveal a community. This example was inspired by similar examples in \citet{arroyo2021inference} and \citet{jones2020multilayer} in the context of multiple graph embedding.
    Consider a tripartite stochastic block model, with two sub-communities of each type, where the matrix $\+B$ has full rank and the form
\begin{equation}
    \label{eq:B_matrix}
    \+B = \left( \begin{array}{c c | c c | c c}
    0 \, & \, 0 \, & \, a \, & \, a \, & \, c \, & \, d \\
    0 \, & \, 0 \, & \, b \, & \, b \, & \, c \, & \, d \\
    \hline
    a \, & \, b \, & \, 0 \, & \, 0 \, & \, e \, & \, e \\
    a \, & \, b \, & \, 0 \, & \, 0 \, & \, f \, & \, f \\
    \hline
    c \, & \, c \, & \, e \, & \, f \, & \, 0 \, & \, 0 \\
    d \, & \, d \, & \, e \, & \, f \, & \, 0 \, & \, 0
    \end{array} \right)
\end{equation}
for some $a,b,c,d,e,f \in [0,1]$.
If, for example, we consider only the bipartite subgraph corresponding to types 1 and 2, the two communities of type 2 are indistinguishable, and in fact, every other bipartite subgraph also obscures a community. No single biadjacency spectral embedding can therefore uncover all of the latent communities. However, they are all revealed through multipartite spectral embedding. This is illustrated by simulation in Figure~\ref{fig:obscured_sbm}.

% \FloatBarrier

\begin{table*}[!p]
    \label{table:drug_clusters}
    \footnotesize
    
    \vspace{-1em}
    \centering
    \caption{Example clusters of drugs.}
    % \hfill
    \begin{tabular}{p{21em}}
    \emph{Cluster 2} \\
    \midrule
    Caffeine \textbf{(Ap, Ce, P1, Ph)}\\
    Theophylline \textbf{(Br, Mu, P1, Ph, Va)}\\
    Adenosine monophosphate \textbf{(Di, Mi, Su)} \\ 
    Aminophylline \textbf{(Br, Ca, Mu, P1, Ph)}\\
    Oxtriphylline \textbf{(Br)}\\
    Flavoxate \textbf{(Pa)} \\
    Papaverine \\
    S,S-(2-Hydroxyethyl)Thiocysteine \\
    Dyphylline \textbf{(Br, Mu, Ph, Va)} \\
    Cilostazol \textbf{(Br, Fi, Np, P3, Pl, Va))} \\
    Guanosine-5\'-Monophosphate \\
    Ketotifen \textbf{(Al, Hi, Pr)} \\
    4-[3-(Cyclopentyloxy)-4-Methoxyphenyl$\ldots$ \\
    % Oxtriphylline \textbf{(Br)} \\
    % Flavoxate \textbf{(Pa)}\\
    $\cdots + 17$ more (30 total)\\
    \midrule
    \textbf{(Va)} 8/35, \textbf{(Br)} 7/22, \textbf{(Ph)} 7/8, \\ \textbf{(Pl)} 5/15, \textbf{(Mu)} 3/19, \textbf{(P1)} 3/3 \\
    \bottomrule
    \end{tabular}
    \hspace{2em}
    \begin{tabular}{p{21em}}
    \emph{Cluster 10} \\
    \midrule
    Pyridoxal Phosphate \textbf{(Di, Mi, Su, Vb)} \\
    Citric Acid \textbf{(Ac, Ch)} \\
    Alglucosidase alfa \textbf{(Ez)} \\
    L-Aspartic Acid \textbf{(Di, Mi, Nea, Su)} \\
    Collagenase \\
    Pyruvic acid \textbf{(Di, Mi, Su)} \\
    Tetrahydrofolic acid \textbf{(Di, Mi, Su)} \\
    Hyaluronan \textbf{(Ad, Vi)} \\
    S-Adenosylmethionine \textbf{(Di, Mi, Su)} \\
    N'-Pyridoxyl-Lysine-5'-Monophosphate \\
    L-Alanine \textbf{(Di, Mi, Nea, Su)} \\
    L-Serine \textbf{(Di, Mi, Nea, Su)} \\
    Salmon Calcitonin \textbf{(Bd, Hc, Os)} \\
    $\cdots + 17$ more (30 total) \\
    \midrule
    \textbf{(Mi)}~10/41,
    \textbf{(Su)}~10/44,
    \textbf{(Di)}~10/44, \\
    \textbf{(Nea)}~3/12, 
    \textbf{(Aa)}~3/6 \\
    % \textbf{(Vb)}~2/10 \\
    \bottomrule
    \end{tabular}
    
    \vspace{2em}
    \begin{tabular}{p{21em}}
    \emph{Cluster 27} \\
    \midrule
    Phenobarbital \textbf{(Cv, Ex, Ga, Hy)}\\
    Diazepam \textbf{(Ad, Am, Ane, Ax, Cv, Ga, Hy, Mu)}\\
    Midazolam \textbf{(Ad, Ane, Ax, Ga, Hy)} \\
    Clobazam \textbf{(Bz, Cv)} \\
    Methylphenobarbital \textbf{(Cv, Ga, Hy)} \\
    Propofol \textbf{(An, Hy)}\\
    Hexobarbital \textbf{(Ad, Ba, Ga, Hy)}\\
    Lorazepam \textbf{(Bz, Hy)}\\
    Triazolam \textbf{(Ad, Ax, Bz, Ga)} \\
    Pentobarbital \textbf{(Ad, Ba, Ga, Hy)} \\
    Secobarbital \textbf{(Ad, Ba, Ga, Hy)} \\
    % Nitrazepam \textbf{(Ax, Cv, Ga, Hy)}\\
    % Alprazolam \textbf{(Bz, Ga, Ax, Hy)} \\
    % Zolpidem \\
    % Thiopental \\
    % Flurazepam \\
    $\cdots + 45$ more (66 total) \\
    \midrule
    \textbf{(Hy)} 29/37, 
    \textbf{(Ga)} 17/18,
    \textbf{(Bz)} 16/17,  \\ \textbf{(Ax)} 12/17, 
    \textbf{(Cv)} 8/32,
    \textbf{(Ad)} 7/25, \\
    \textbf{(Ba)} 6/7,
    \textbf{(Ane)} 5/31 \\
    \bottomrule
    \end{tabular}    
    \hspace{2em}
    \begin{tabular}{p{21em}}
    \emph{Cluster 32} \\
    \midrule
    Methadone \textbf{(An, Na, Tu)}\\
    Morphine \textbf{(An, Na)}\\
    Heroin \textbf{(An, Na)}\\
    Oxycodone \textbf{(Ad, An, Na)}\\
    Fentanyl \textbf{(Ad, An, Ana, Na)} \\
    Ketamine \textbf{(Ag, Ane, Ex)}\\
    Alfentanil \textbf{(Ag, An, Na)} \\
    Pethidine \textbf{(Ad))} \\
    Loperamide \textbf{(Dr)} \\
    Ketobemidone \textbf{(Ana, Ex)} \\
    Hydrocodone \textbf{(Ana, Na, Tu)} \\
    Levomethadyl Acetate \textbf{(Ana, Na)} \\
    Hydromorphone \textbf{(Ana, Na)} \\
    $\cdots + 17$ more (30 total) \\
    \midrule
    \textbf{(Ag)} 20/41, \textbf{(Na)} 17/18,  \textbf{(Ad)} 5/25, \\ \textbf{(An)} 5/31,
    \textbf{(Naa)} 5/8,
    \textbf{(Tu)} 5/8 \\
    \bottomrule
    \end{tabular}
    % \hfill
    
    \vspace{2em}
    % \addtabletext{
    \begin{center}``\textbf{{(Xx)}} $a/b$'' means ``label \textbf{{(Xx)}} appears $a$ times in the cluster and $b$ times in total''.\end{center}
    \vspace{1em}
    Labels:
    \textbf{{(Aa)}} Amino acids,
    \textbf{{(Ad)}} Adjuvants,
    \textbf{(Al)} Anti allergic agents
    \textbf{{(Ana)}} Analgesics,
    \textbf{{(Ane)}} Anesthetics,
    \textbf{{(Ap)}} Appetite depressants,
    \textbf{{(Ax)}} Anti-anxiety agents,
    \textbf{(Ba)} Barbiturates,
    \textbf{{(Bd)}} Bone density conservation agents,
    \textbf{{(Br)}} Bronchodilator agents,
    \textbf{{(Bz)}} Benzodiazepines,
    \textbf{{(Ca)}} Cardiotonic agents,
    \textbf{{(Ce)}} Central-nervous-system stimulants,
    \textbf{{(Cv)}} Anticonvulsants,
    \textbf{{(Di)}} Dietary supplements,
    \textbf{(Dr)} Antidiarrheals,
    \textbf{{(Ex)}} Excitatory amino acid antagonists,
    \textbf{{(Ez)}} Enzyme replacement agents,
    \textbf{(Fi)} Fibrinolytic agents
    \textbf{{(Ga)}} GABA modulators,
    \textbf{{(Hc)}} Antihypercalcemic agents,
    \textbf{(Hi)} Histamine H1 antagonists
    \textbf{{(Hy)}} Hypnotics and sedatives,
    \textbf{{(Mi)}} Micronutrients,
    \textbf{{(Mu)}} Muscle relaxants,
    \textbf{{(Na)}} Narcotics,
    \textbf{{(Naa)}} Narcotic antagonists,
    \textbf{{(Nea)}} Non-essential amino acids,
    \textbf{{(Nm)}} Neuromuscular agents,
    \textbf{(Np)} Neuroprotective agents
    \textbf{{(Os)}} Anti-osteportic agents,
    \textbf{{(P1)}} Purinergic P1 receptor antagonists,
    \textbf{(P3)} Phosphodiesterase-3 inhibitors
    \textbf{(Pa)} Parasympatholytics,
    \textbf{{(Ph)}} Phosphodiesterase inhibitors,
    \textbf{(Pl)} Platelet aggregation inhibitors,
    \textbf{(Pr)} Antipruritics,
    \textbf{{(Su)}} Supplements,
    \textbf{{(Tu)}} Antitussive agents,
    \textbf{{(Va)}} Vasodilator agents,
    \textbf{{(Vb)}} Vitamin-B complex,
    \textbf{(Vi)} Viscosupplements.
    % } 
    \end{table*}
    
    % \vspace{-1em}
    \begin{table*}[!p]
    \label{table:pathway_clusters}
    \footnotesize
    \centering
    \caption{Example clusters of pathways.}
    \begin{tabular}{p{21em}}
    \emph{Cluster 26} \\
    \midrule
    Valine, leucine and isoleucine degradation \textbf{(Am, Me)} \\
    Fatty acid degradation \textbf{(Li, Me)} \\
    Pyruvate metabolism \textbf{(Cr, Me))} \\
    Propanoate metabolism \textbf{(Cr, Me))} \\
    Terpenoid backbone biosynthesis \textbf{(Me, Tp)} \\
    Fatty acid metabolism \textbf{(Me, Ov)} \\
    Glycerolipid metabolism \textbf{(Li, Me)} \\
    Fatty acid biosynthesis \textbf{(Li, Me)} \\
    \midrule
    \textbf{(Me)} 8/80, 
    \textbf{(Li)} 3/15,
    \textbf{(Or)} 2/69 \\
    \bottomrule
    \end{tabular}
    \hspace{2em}
    \begin{tabular}{p{21em}}
    \emph{Cluster 33} \\
    \midrule
    Alzheimer's disease \textbf{(Hu, Ne)}\\
    Parkinson's disease \textbf{(Hu, Ne)}\\ 
    Oxidative phosphorylation \textbf{(Em, Me)} \\
    Non-alcoholic fatty liver disease \textbf{(Em, Hu)} \\
    Huntington's disease \textbf{(Hu, Ne)} \\  \\ \\ \\ \\
    \midrule
    \textbf{(Hu)} 4/71, \textbf{(Ne)} 3/5 \\
    \bottomrule
    \end{tabular}
    
    \vspace{2em}
    \begin{tabular}{p{21em}}
    \emph{Cluster 37} \\
    \midrule
    Prostate cancer \textbf{(Ca, Hu)}\\
    Central carbon metabolism in cancer \textbf{(Ca, Hu)}\\
    Chronic myeloid leukemia \textbf{(Ca, Hu)}\\
    Melanoma \textbf{(Ca, Hu)} \\
    Non-small cell lung cancer \textbf{(Ca, Hu)}\\
    Glioma \textbf{(Ca, Hu)} \\
    Renal cell carcinoma \textbf{(Ca, Hu))}\\
    Acute myeloid leukemia \textbf{(Ca, Hu)} \\
    Bladder cancer \textbf{(Ca, Hu)} \\
    Thyroid cancer \textbf{(Ca, Hu)} \\
    Dorso-ventral axis formation \textbf{(De, Or)}\\ \\ \\ \\ \\
    \midrule
    \textbf{(Ca)} 10/22, \textbf{(Hu)} 10/71 \\
    \bottomrule
    \end{tabular}
    \hspace{2em}
    \begin{tabular}{p{21em}}
    \emph{Cluster 41} \\
    \midrule
    cAMP signaling pathway \textbf{(Ei, Si)} \\
    Morphine addiction \textbf{(Hu, Su)}\\
    Alcoholism \textbf{(Hu, Su)} \\
    Amphetamine addiction \textbf{(Hu, Su)}\\
    Circadian entrainment \textbf{(En, Or)}\\
    Long-term potentiation \textbf{(Nr, Or)} \\
    Amyotrophic lateral sclerosis (ALS) \textbf{(Hu, Ne)} \\
    Nicotine addiction \textbf{(Hu, Su)}\\
    Renin secretion \textbf{(En, Or)}\\
    Cocaine addiction \textbf{(Hu, Su)}\\
    Thyroid hormone synthesis \textbf{(En, Or)} \\
    Taste transduction \textbf{(Or, Se)} \\
    Vibrio cholerae infection \textbf{(In, Hu)} \\
    Olfactory transduction \textbf{(Se, Or)} \\
    \midrule
    \textbf{(Hu)} 7/71, \textbf{(Or)} 6/69, \textbf{(Su)} 5/5, \\ \textbf{(En)} 2/17, \textbf{(Se)} 2/4 \\
    \bottomrule
    \end{tabular}
    
    \vspace{2em}
    % \addtabletext{
    \begin{center}``\textbf{{(Xx)}} $a/b$'' means ``label \textbf{{(Xx)}} appears $a$ times in the cluster and $b$ times in total''.\end{center}
    % \vspace{-.6em}
    \vspace{1em}
    Labels:
    \textbf{(Am)} Amino acis metabolism,
    \textbf{{(Ca)}} Cancers,
    \textbf{{(Ce)}} Cellular processes,
    \textbf{(Cr)} Carbohydrate metabolism,
    \textbf{(De)} Development,
    \textbf{{(Di)}} Digestive system,
    \textbf{(Ei)} Environmental information processing,
    \textbf{{(Em)}} Energy metabolism,
    \textbf{{(En)}} Endocrine systems,
    \textbf{{(Hu)}} Human diseases,
    \textbf{(In)} Infectious diseases
    \textbf{{(Li)}} Lipid metabolism,
    \textbf{{(Me)}} Metabolism,
    \textbf{{(Ne)}} Neurodegenerative diseases,
    \textbf{(Nr)} Nervous system,
    \textbf{{(Or)}} Organismal systems,
    \textbf{{(Ov)}} Overview,
    \textbf{(Se)} Sensory system,
    \textbf{(Si)} Signal transduction, 
    \textbf{{(Su)}} Substance dependence,
    \textbf{(Tp)} Metabolism of terpenoids and polyketides, 
    \textbf{{(Tr)}} Transport and catabolism.
    % }
    
    \end{table*}

\section{Real data}
\label{sec:real_data}

Here we apply Algorithm~\ref{alg:clustering} to a multipartite network representing associations between biomedical entities of six distinct types (Figure~\ref{fig:biomedical_network_diagram}): drugs, diseases, targets, pathways, variant locations and haplotypes. Figure~\ref{fig:biomedical_network_diagram} shows a schematic of the topology of the data. The associations were inferred from several biological databases: Drugbank \citep{wishart2006drugbank}, Kyoto Encyclopedia of Genes and Genomes (KEGG) \citep{kanehisa2000kegg}, PharmGKB \citep{hewett2002pharmgkb} and the Human Disease network \citep{goh2007human}. A superset of the dataset we use was introduced and detailed in \citet{zong2021drug}.

% \citep{wishart2006drugbank,belleau2008bio2rdf,goh2007human,kanehisa2000kegg, hewett2002pharmgkb}, and 
 
We apply Algorithm~\ref{alg:clustering} to the graph,  implementing Laplacian normalisation as described in Remark~\ref{remark:adj_rescaling} with $\tau$ equal the average degree of the graph as recommended in \citet{qin2013regularized}. We use the elbow method of \citet{zhu2006automatic} for ambient and intrinsic dimension selection, implement the optional spherical projection step, and select the number of communities in each block equal to the intrinsic dimension estimates under the assumption of a full-rank degree-corrected stochastic block model. In addition, we do not include nodes of degree less than five in the clustering steps, on the basis that there is not enough signal in their positions for them to be accurately clustered.

% Algorithm~?????????????????????????????????????????????????????????????????????????????????????????????????????????????????????????????????????????????????????????????????????????????????????????????????????????????????????????????????????????????????????????????????????????????????????????????????????????????????????????????????????????????????????????????????????????????????\ref{alg:clustering} will be applied in the following configuration: we implement the optional spherical projection step (line 6); we use a regularized variant of the normalized Laplacian matrix \citep{chaudhuri2012spectral, amini2013pseudo} as input, in which the degree matrix is inflated to D+τID + \tau I, with τ\tau set to the average degree, as recommended in \citet{qin2013regularized}; we use the method of \citet{zhu2006automatic} for rank selection (lines 1 and 4);

% finally, at the clustering step (line 7), we remove nodes with degree less than five. 

The left panel of Figure~\ref{fig:scree_plots} shows the first 1000 singular values of the regularized Laplacian and the dimension ($\hat r=214$) selected by the elbow method of \citet{zhu2006automatic}. The remaining panels show the singular values of the ambient embedding of each node type. The black lines show the intrinsic dimension selected by the elbow method and the dashed line shows $\min\{\hat p,\hat q\}$, which is always larger, as predicted by Lemma~\ref{lemma:subspacec_dimension}.

The intrinsic dimension selected also acts as an estimate of the number of communities of that type.
% So, in the first group, Drugs, we apply kk-means with 6161 clusters.
Drug and Pathway nodes each have associated labels, obtained from the DrugBank \citep{wishart2006drugbank} and KEGG \citep{kanehisa2000kegg} databases, respectively, which we use as held-out information to interpret and evaluate the clustering obtained. 

The correspondence between the communities recovered and their labels is strong. %(xx reword).  
Tables 1 and 2 show example clusters of Drugs and Pathways. The items with the highest degree are shown, with their labels in brackets. Below, we show the number of occurrences of each label within the cluster and in total. We omit labels which occur twice or less in the Drug clusters and which occur once or less in the Pathways clusters.

In the Drug clusters, Cluster 2 contains primarily chemicals which dilate the blood vessels or airways, Cluster 10 contains primarily nutrition-related substances, Cluster 27 contains primarily hypnotics and sedatives including all but one of the benzodiazepines and GABA modulators, and Cluster 61 includes all but one of the narcotics. In the Pathway clusters, Cluster 26 corresponds to pathways related to metabolism, Cluster 39 to neuro-degenerative diseases, Cluster 37 to cancer and Cluster 41 to substance dependence.

% Code to reproduce the analysis in this section is available at {\texttt{\small{github.com/alexandermodell/multipartite\_clustering}}}.
% Any reasonable test against the null hypothesis that the labels are uniformly distributed among clusters gives a p-value close to zero. 

\section*{Acknowledgement}
The authors thank Benjamin Barrett for useful discussions and Nansu Zong for providing the data.

% \section*{Disclosure statement}
% The authors report there are no competing interests to declare.

\bibliographystyle{apalike}
\bibliography{multipartite}

\newpage
\appendix

\appendix
\section*{Appendix}

\section{Proof of Lemma~\ref{lemma:subspacec_dimension}}
\label{sec:mp_subspace_dim_proof}
Let $\*V$ be a totally isotropic subspace with respect to the indefinite inner product $\langle x, y \rangle_{p,q} = x^\top\Ipq y$, so that $\langle x, y \rangle_{p,q} = 0$ for any $x, y \in \*V$.
Let $\*W$ be an arbitrary subspace of dimension $\max\{p,q\}$ which is either positive or negative definite, so that $\langle x, y \rangle_{p,q} = 0$ implies $x=y=0$, for any $x,y\in \*W$. Then $\*V \cap \*W = \{0\}$, so $\dim(\*V+\*W) = \dim(\*V)+\dim(\*W)$ and
  \begin{equation*}
    \dim(\*V) = \dim(\*V + \*W)  - \dim(\*W) \leq p+ q - \max\{p,q\} = \min\{p,q\}.
  \end{equation*}
Therefore, the maximal dimension of a totally isotropic subspace with respect to the indefinite inner product $\langle \cdot, \cdot \rangle_{p,q}$ is $\min \{p,q\}$.
% Recall that $P = X\Ipq X^\top$ and that by definition, the rows of $X$ corresponding to a single partition have indefinite inner product zero and live in a totally isotropic subspace. Therefore, the result follows.

\section{Proof of Theorem~\ref{thm:xie_consistency}}

In this section, we show that Theorem~1 follows from Corollary~3.1 of \citet{xie2021entrywise} under the stated conditions. The Corollary in \citet{xie2021entrywise} is stated under a random dot product graph model, however a simple adaptation of the proof shows that the bound holds under a generalised random dot product graph model, up to a multiplicative factor of $\smallnorm{Q_X}$, where $Q_X \in \O(p,q)$ is the matrix such that $X = U|S|^{1/2}Q_X$, where $P = USU^\top$ denotes a singular value decomposition of $P$, with $U\in\R^{n\times r}$ and $S \in \R^{r\times r}$. The following lemma, the proof of which we defer to the end of the section, provides control on the spectral norm of $Q_X$ and $Q_X^{-1}$. 

\begin{lemma}
\label{lemma:Q_norm}
    Let $P = USU^\top$ denote a singular value decomposition of $P$, with $U\in\R^{n\times r}$ and $S \in \R^{r\times r}$, and let $Q_X \in \O(p,q)$ be the matrix such that $X = U|S|^{1/2}Q_X^\top$. Then
    \begin{equation*}
        \norm{Q_X}_2, \norm{Q_X^{-1}}_2 \leq \kappa^{1/2}(\Delta).
    \end{equation*}
\end{lemma}

By Lemma~\ref{lemma:Q_norm}, and the assumption that $\kappa(\Delta) \eqc 1$, we have that $\smallnorm{Q_X} \leqc 1$. In our notation, Corollary~3.1 of \citet{xie2021entrywise} then states that with overwhelming probability, there exists $Q \in \O(p,q)$ such that
\begin{equation}
\label{eq:xie_full_bound}
    \max_{i \in [n]} \norm{Q \hat X_i - X_i}_2 \leqc \frac{\rho^{3/2}\norm{U}_{2,\infty}}{n^{1/2} \lambda_d^2(\Delta)} \max \curlybrackets{\frac{\rho \log^{1/2}n}{\lambda_d(\Delta)}, \frac{\rho^2}{\lambda_d^2(\Delta)}, \log n} + \frac{\norm{U}_{2,\infty}(\rho \log n)^{1/2}}{\lambda_d^{1/2}(\Delta)}.
\end{equation}
We first observe that since $\norm{X_i} \eqc \rho^{1/2}$ for all $i \in [n]$, $\norm{X}_{\text{F}}^2 \eqc n \rho$ and therefore
    \begin{equation*}
        \lambda_1(\Delta) = n^{-1}\norm{X^\top X}_2 = n^{-1} \norm{X}_2^2 \geq \frac{1}{n}\norm{X}_{\text{F}} \eqc \frac{n\rho}{n} = \rho.
    \end{equation*}
By assumption, $\lambda_1(\Delta) \eqc \lambda_d(\Delta)$, and therefore $\lambda_d(\Delta) \eqc \rho$. In addition,
\begin{align*}
    \norm{U}_{2,\infty} &\leq \norm{U|S|^{1/2}}_{2,\infty} \norm{|S|^{-1/2}}_2 \\
    &= \norm{XQ^{-1}}_{2,\infty} \norm{|S|^{-1/2}}_2 \\
    &\leq \norm{X}_{2,\infty} \norm{Q^{-1}}_2 \norm{|S|^{-1/2}}_2 \\ 
    &\leqc \brackets{\frac{ \rho \kappa(\Delta)}{n\lambda_d(\Delta)}}^{1/2} \\
    &\eqc n^{-1/2}
\end{align*}
where the final equality follows from the assumption that $\kappa(\Delta) \eqc 1$. Plugging these quantities into \eqref{eq:xie_full_bound}, simplifying and employing the assumption that $n \rho \geqc \log n$, we obtain Theorem~\ref{thm:xie_consistency}.

\section{Proof of Lemma~\ref{lemma:Q_norm}}
To prove Lemma~\ref{lemma:Q_norm}, we first observe that
\begin{equation*}
    Q|S|Q^\top = (U|S|^{1/2}Q^\top)^\top U|S|^{1/2}Q = X^\top X = n\Delta
\end{equation*}
and by an elementary min-max argument, we have that
\begin{equation*}
n\lambda_1\brackets{\Delta} = \lambda_1\brackets{Q|S|Q^\top} \geq \lambda_d \brackets{|S|}\lambda_1^2\brackets{Q} = n\lambda_d \brackets{\Delta}\lambda_1^2\brackets{Q}
\end{equation*}
where the last equality since $|S|$ and $n\Delta$ have the same eigenvalues. Rearranging yields
\begin{equation*}
    \lambda_1\brackets{Q} \leq \brackets{\frac{\lambda_1(\Delta)}{\lambda_d(\Delta)}}^{1/2} = \kappa^{1/2}(\Delta),
\end{equation*}
which establishes that $\norm{Q}_2 \leq \kappa^{1/2}(\Delta)$. To establish the second claim, we observe that $Q^{-1} = \Ipq Q \Ipq$, and therefore that
\begin{equation*}
    \norm{Q^{-1}}_2 = \norm{\Ipq Q \Ipq}_2 \leq \norm{\Ipq}_2\norm{Q}_2\norm{\Ipq}_2 = \norm{Q}_2 \leq \kappa^{1/2}(\Delta)
\end{equation*}
since $\norm{\Ipq}_2 = 1$.
    
\section{Proof of Theorem~\ref{thm:mp_consistency}}
\label{sec:mp_proof}

% We begin by defining some notation. We say an\leqcPbna_n \leqcP b_n to mean that an\leqcbna_n \leqc b_n with overwhelming probability. We use ‖\|\cdot\|_2 to denote the spectral norm, and \|\cdot\|_{2\to\infty}\|\cdot\|_{2\to\infty} denote the two-to-infinity norm, namely, the maximum of the row-wise Euclidean norms \citep{cape2019two}. Let \O(d)\O(d) and \GL(d)\GL(d) denote the dd-dimensional orthogonal and general linear groups, respectively, and let \O(p,q) = \{M \in \R^{d\times d} : MI_{p,q}M^\top = I_{p,q}\}\O(p,q) = \{M \in \R^{d\times d} : MI_{p,q}M^\top = I_{p,q}\} be the indefinite orthogonal group with signature (p,q)(p,q).

% We say $a_n \leqcP b_n$ to mean that $a_n \leqc b_n$ with overwhelming probability.
In this proof, we use the notation $a_n \leqcP b_n$ to denote that the event $a_n \leqc b_n$ holds with overwhelming probability.

The first step of our proof is to define vectors $X_1,\ldots,X_n \in \R^r$ such that $Y_i^\top \Lambda_{z_i,z_j} Y_j = \inner{X_i, X_j}_{p,q}$ for all $i,j \in [n]$, so that $\+A$ is described as a generalised random dot product graph, and we can employ existing estimation theory in \citet{rubin2022statistical}.

We construct the $(r_1 + \cdots + r_K) \times (r_1 + \cdots + r_K)$ block-matrix $\+{\Lambda}$ whose $k\ell$th block is $\+{\Lambda}_{k\ell}$, and recall the matrices $\+{\Lambda}_k = (\+{\Lambda}_{k1} \: \cdots \: \+{\Lambda}_{kK})$, the row concatenation of $\+{\Lambda}_{k1},\ldots,\+{\Lambda}_{kK}$. Let $r, (p,q)$ denote the rank and signature, respectively, of $\+{\Lambda}$ and let $\+H$ be an $r \times (r_1+\cdots+r_K)$ matrix such that $\+{\Lambda} = \+H^\top \Ipq \+H$, which can be constructed as $\+H = \+U_{\+{\Lambda}} |\+S_{\+{\Lambda}}|^{1/2}$, where $\+{\Lambda} = \+U_{\+{\Lambda}}\+S_{\+{\Lambda}}\+U_{\+{\Lambda}}^\top$ is the eigendecomposition of $\+{\Lambda}$. In addition, let $\+H_k$ denote the $k$th block of $\+H$ containing the rows of $\+H$ corresponding to the $k$th group. Now, by construction, $\+{\Lambda}_{k\ell} = \+H_k^\top \Ipq \+H_\ell$ and $\+{\Lambda}_k = \+{H}_k^\top \Ipq \+{H}$. Since by assumption, $\+{\Lambda}_k$ has rank $d_k$, so does $\+H_k$. We construct $X_i = \+H_{z_i}Y_i$ which satisfies
\begin{align*}
    Y_i^\top \+{\Lambda}_{z_i,z_j} Y_j = Y_i^\top \+H_{z_i}^\top \Ipq \+H_{z_j} Y_j = X_i^\top \Ipq X_j = \inner{X_i,X_j}_{p,q},
\end{align*}
as desired.

Observe that since the transformations $\{\+H_k\}_{k=1}^K$ are fixed, $\smallnorm{X_i}_2 \eqc \smallnorm{Y_i}_2 \eqc \rho^{1/2}$ and from the assumptions that $\kappa(\+\Lambda) \eqc 1$, it is straight-forward to derive that $\kappa(\+\Delta) := \sigma_1(\+\Delta) / \sigma_r(\+\Delta) \eqc 1$, where $\+\Delta := n^{-1}\sum_{i=1}^n X_i X_i^\top$. Therefore the assumptions of Theorem~\ref{thm:xie_consistency} are satisfied, and we have that there exists an indefinite orthogonal matrix $\+Q \in \O(p,q)$ such that
% \begin{equation}
% \label{eq:mp_grdpg_tti_Q}
%     \max_{i \in [n]} \norm{\+Q^{-1}\hat X_i -  X_i}_2 \leqcP \brackets{\frac{r\log n}{n}}^{1/2}.
% \end{equation}
\begin{equation}
\label{eq:mp_grdpg_tti}
    \max_{i \in [n]} \norm{\hat X_i - \+Q X_i}_2 \leqcP \brackets{\frac{\log n}{n}}^{1/2}.
\end{equation}
By Lemma~\ref{lemma:Q_norm}, $\smallnorm{Q}_2, \smallnorm{Q^{-1}}_2 \leq \kappa^{1/2}(\Delta) \eqc 1$.
It therefore follows that
\begin{align*}
    \max_{i\in[n]}\norm{\hat X_i}_2 &\leq \max_{i\in[n]} \norm{\+Q X_i + \hat X_i - \+Q X_i}_2 \\
    &\leq \norm{\+Q}_2\max_{i\in[n]}\norm{X_i}_2 + \max_{i\in[n]}\norm{\hat X_i - \+Q X_i}_2 \\
    &\leqcP \rho^{1/2} + \brackets{\frac{\log n}{n}}^{1/2} \\
    &\eqc \rho^{1/2}.
\end{align*}
where we used that $n \rho_n \geqc \log n$. Now recall the matrix $\hat{\+\Sigma}_k = n_k^{-1} \sum_{i \in V_k} \hat X_i \hat X_i^\top$ and that $\hat{\+\Xi}_k$ is the matrix whose columns contain the $r_k$ orthonormal eigenvectors of $\hat{\+\Sigma}_k$ corresponding to the largest eigenvalues. We define its population counterpart $\+\Sigma_k = n_k^{-1} \sum_{i \in V_k} X_i X_i^\top$ and define the matrix $\+\Xi_k$ whose columns contain the $r_k$ eigenvectors of $\+Q \+\Sigma_k \+Q^\top$ corresponding its non-zero eigenvalues.
We have that
\begin{align*}
    \norm{\hat{\+\Sigma} - \+Q\+\Sigma \+Q^\top}_2 &= \norm{n_k^{-1}\sum_{i \in  V_k} \brackets{\hat X_i \hat X_i^\top - \+Q X_i X_i^\top \+Q^\top}}_2 \\
    &= \norm{n_k^{-1}\sum_{i\in V_k} \curlybrackets{\hat X_i \brackets{\hat X_i - \+Q X_i}^\top + \brackets{\hat X_i - \+Q X_i}X_i^\top \+Q^\top}}_2 \\
    &\leq \brackets{\max_{i\in V_k} \smallnorm{\hat X_i}_2 + \max_{i\in V_k} \smallnorm{X_i}_2 \smallnorm{\+Q}_2} \cdot n_k^{-1} \sum_{i\in V_k} \norm{\hat X_i - \+Q X_i}_2 \\
    &\leqcP \brackets{\frac{\rho \log n}{n}}^{1/2}.
\end{align*}
The smallest non-zero eigenvalue of $\+Q \+{\Sigma}_k \+Q^\top$, $\delta_k$ satisfies $\delta_k \eqc \rho$ and by the Davis-Kahan sin$\Theta$ theorem, we have that there exists an orthogonal matrix $\+W_k \in \O(d_k)$ such that

\begin{equation}
\label{eq:mp_dk}
    \norm{\hat{\+\Xi}_k - \+\Xi_k \+W_k}_2 \leqc \delta_k^{-1} \norm{\hat{\+\Sigma}_k - \+Q \+\Sigma_k \+Q^\top}_2 \leqcP \brackets{\frac{\log n}{n\rho}}^{1/2}.
\end{equation}
We set $\+G_k := \+W_k^\top \+\Xi_k^\top \+Q (\+H_k \+H_k^\top)^{-1} \+H_k$ and then we have
\begin{align*}
    \hat Y_i - \+G_{z_i} Y_i &= \hat{\+\Xi}_{z_i}^\top \hat X_i - \+G_{z_i} \+H_{z_i}^\top X_i \\
    &= \hat{\+\Xi}_{z_i}^\top \hat X_i - \+W_{z_i}^\top \+\Xi_{z_i}^\top \+Q X_i \\
    &= \hat{\+\Xi}_{z_i}^\top \hat X_i - \hat{\+\Xi}_{z_i}^\top \+Q X_i + \hat{\+{\Xi}}_{z_i}^\top \+Q X_i - \+W_{z_i}^\top \+{\Xi}_{z_i}^\top \+Q X_i \\
    &= \hat{\+\Xi}_{z_i}^\top \brackets{\hat X_i - \+Q X_i} + \brackets{\hat{\+\Xi}_{z_i} - \+W_{z_i} \+{\Xi}_{z_i}}^\top \+Q X_i
\end{align*}
Therefore it follows from \eqref{eq:mp_grdpg_tti}, \eqref{eq:mp_dk} and the bounded spectral norm of $\+Q$, that
\begin{align*}
    \max_{i \in [n]} \norm{\hat Y_i - \+G_{z_i} Y_i}_2 &\leq \max_{i \in [n]} \norm{\hat X_i - \+Q X_i}_2 + \max_{k \in [K]} \norm{\hat{\+\Xi}_{k} - \+W_{k} \+\Xi_{k}}_2 \norm{\+Q}_2 \max_{i \in [n]} \norm{X_i}_2 \\
    &\leqcP \brackets{\frac{\log n}{n}}^{1/2},
\end{align*}
which establishes the theorem.

\end{document}